# Sensitivity analysis in real options studies: systematic review and recommendations

Mariia Kozlova[a] and Julian Scott Yeomans[b]

[a] LUT Business School, LUT University, Lappeenranta 53850, Finland. Mariia.Kozlova@lut.fi

[b] Schulich School of Business, York University, 4700 Keele Street, Toronto, ON, M3J 1P3 Canada. syeomans@yorku.ca

## Abstract

Real options (RO) modeling provides a powerful framework for valuing managerial flexibility under uncertainty in capital-intensive investments across various industries, including climate, energy, and transportation. However, the transparency and usefulness of such models depends heavily on sensitivity analysis (SA). This study reviews over 100 peer-reviewed RO journal articles to examine the extent and quality of SA practices. The review focuses particularly on the extent of model scrutiny: how many input variables are subjected to sensitivity analysis; whether local or global approaches are applied; and whether the shape of the input–output relationships is examined. Our findings reveal a strong overreliance on one-at-a-time (OAT) techniques and a corresponding minimal adoption of global sensitivity analysis (GSA). This methodological gap limits the ability of the RO field to critically examine its own models and to support its decision-making. The paper concludes by outlining concrete recommendations for how to effectively incorporate GSA into future RO studies in order to enhance model transparency, relevance, and credibility.

## 1. Introduction

Despite the widespread reliance on deterministic models in decision-making, there is growing recognition that deriving single-valued outputs from such models may prove to obscure the underlying uncertainties and mislead the ensuing policy and investment choices (Saltelli et al., 2020). While sensitivity analysis (SA) is essential to probe model behavior under uncertainty and the influence of methodological choices on the results (Hämäläinen & Lahtinen, 2016), in many fields it is commonly reduced to one-at-a-time (OAT) perturbations of the inputs (Pianosi et al., 2016, Lo Piano et al., 2021, Kozlova, Lo Piano, & Yeomans, 2024). OAT approaches capture only isolated directions in the input space and leave the vast majority of potential variability, and all joint input effects, unexplored.

Global sensitivity analysis (GSA) was introduced in the early 1990s as a solution to the limitations of local, derivative-based, and OAT methods (Tarantola et al., 2024). Unlike local SA, GSA quantifies the contribution of each input to the overall output variability across the full range of input uncertainty. Over the past three decades, GSA has evolved significantly, with the continuous development of variance-based methods (Sobol', 1993, Saltelli, 2000, Kucherenko et al., 2019, Kozlova et al., 2025), moment-independent approaches (Borgonovo, 2007; Da Veiga, 2015; Puy et al., 2023) and other branches. This methodological expansion has made GSA applicable to a broad spectrum of model types, including non-linear, non-monotonic, and models with dependent inputs. Yet, despite its wide applicability, GSA remains underutilized in many modeling fields, including real options, where a lack of institutionalized instruction (Saltelli et al., 2019), insufficient structuring, and standardizing SA as a discipline act as persistent barriers to adoption (Razavi et al., 2021).

The field of real options (RO) represents a methodological advance of deterministic cost-benefit analysis by design that values flexibility and incorporates uncertainty into project decision-making (Trigeorgis, 1996). However, despite this conceptual leap, RO models have remained hard-wired to yield deterministic outputs. For example, crisp option value and optimal timing results can lead to non-transparent modeling practices, an overreliance on quantification, and limited decision-making support (Saltelli et al., 2020). A Scopus search of RO publications reveals that sensitivity analysis is rarely featured: the majority of papers do not include SA at all in their abstracts or keywords, while fewer than 1% explicitly refer to global methods, Figure 1.

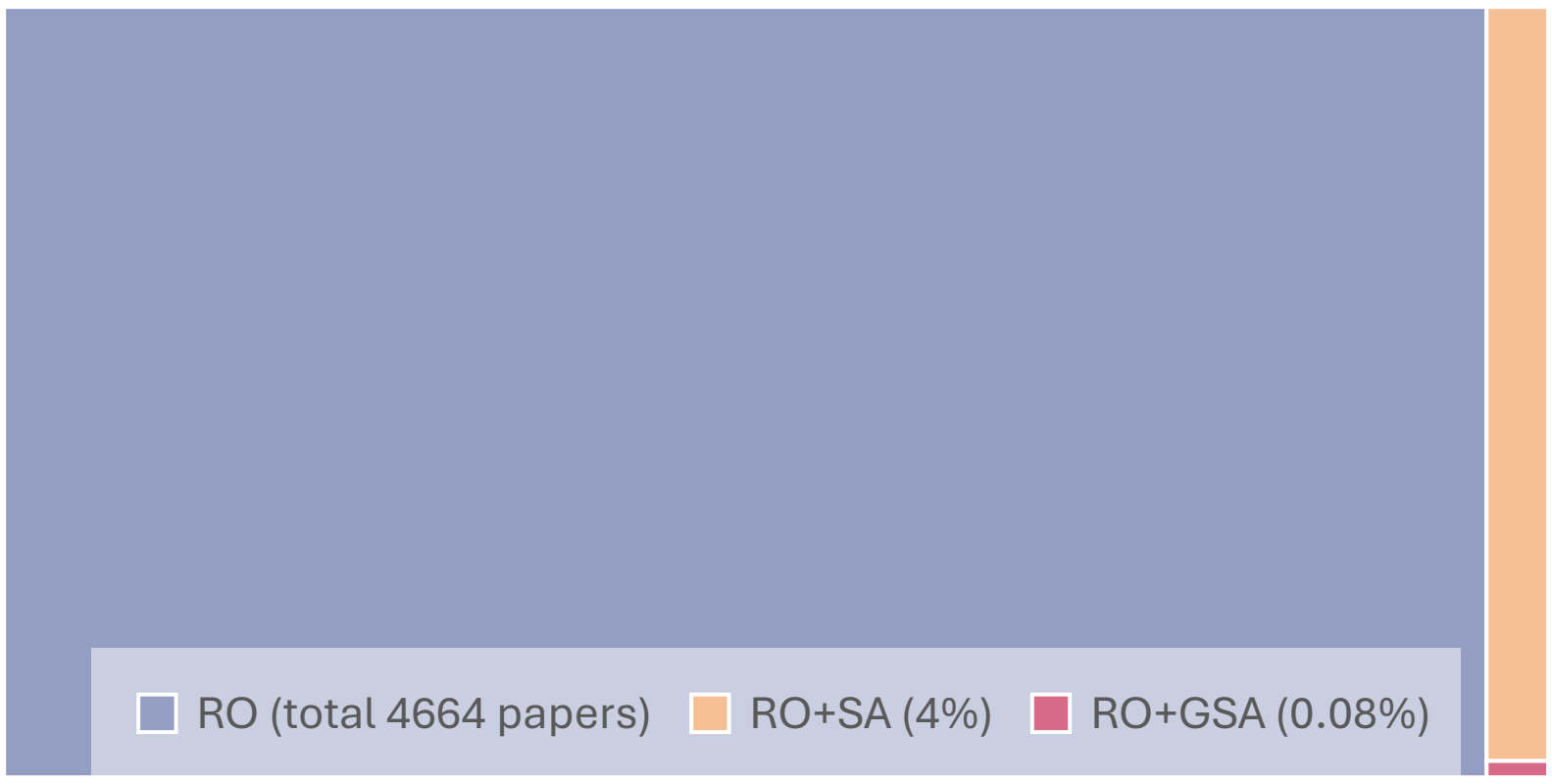

**Figure 1**. Treemap chart showcasing the entity of publications on real options (Scopus) and shares of papers that refer to sensitivity analysis (SA) and to global sensitivity analysis (GSA) in their title, abstract or keywords.

To investigate the issue further, this paper reviews SA practices and confirms the prevalence of oversimplified techniques throughout the RO literature. To counter these observed shortcomings, numerous actionable recommendations are proposed for integrating more rigorous, informative sensitivity analyses into RO workflows.

## 2. Existing reviews and contribution of this paper

There are several previous studies that have reviewed the real options literature, Table 1. Each of these studies has followed a similar blueprint. First, they focus on a specific industrial domain, e.g. energy or transport (except Trigeorgis and Tsekrekos' (2018) who focus on operations research), Second, they analyse dozens of papers in terms of their thematic application and RO study design, including valuation method, uncertainty modeling, and options included. Third, they entirely skip the dimension of sensitivity analysis. From the eight considered review papers, six do not mention sensitivity analysis at all; one (Trigeorgis & Tsekrekos, 2018) quotes 'sensitivity analysis' among key words of other two papers, without opening up the term; and one (Schachter & Mancarella, 2016) mentions its infrequent use, but does not go into any detail.

**Table 1.** Existing systematic reviews of real options literature.

| Paper | Journal* | Number of papers reviewed | Focus | Dimensions of analysis | | |
|---|---|---|---|---|---|---|
| | | | | Theme | RO study design | SA |
| (Nadarajah & Secomandi, 2023) | EJOR | 80 | energy | yes | yes | no |
| (Zheng & Jiang, 2023) | TEM | 51 | transport | yes | yes | no |
| (Lazo & Watts, 2023) | SETA | 92 | PV energy | yes | yes | no |
| (Agaton, 2021) | STA | 67 | carbon capture | yes | yes | no |
| (Trigeorgis & Tsekrekos, 2018) | EJOR | 164 | operations research | yes | yes | no |
| (Kozlova, 2017) | RSER | 101 | renewable energy | yes | yes | no |
| (Schachter & Mancarella, 2016) | RSER | not specified | smart grid | yes | yes | no** |
| (Ceseña et al., 2013) | RSER | 41 | energy | yes | yes | no |
| **This review** | | **131** | **sensitivity analysis** | **yes** | **yes** | **yes** |

* EJOR – European Journal of Operational research, TEM - Transport Economics and Management, SETA - Sustainable Energy Technologies and Assessments, STE - Science of the Total Environment, RSER – Renewable and Sustainable Energy Reviews.

** Schachter and Mancarella (2016) mention that sensitivity analysis is rarely used in the sample of reviewed papers, but they do not specify any further details.

This review addresses this omission by specifically focusing on RO papers that mention sensitivity analysis and scrutinizing the choices behind its methodological design in-depth. Our findings suggest ample room exists for improvement to the practice of sensitivity analysis within the RO community. Drawn recommendations will help transform modeling practices towards better transparency, accountability, and practical usefulness.

## 3. Brief overview of real options

While prevalent in investment valuation, the discounted cash flow (DCF) method has long been criticized because of its deterministic nature and its inability to account for uncertainty and managerial flexibility (Trigeorgis, 1996, Amram & Kulatilaka, 1999). In response to these limitations, the concept of applying financial option logic to real assets gave rise to the field of real options. The philosophical foundation of real options theory is rooted in the recognition that flexibility carries intrinsic value in an uncertain world. Appropriately accounting for flexibility can significantly enhance the attractiveness and strategic resilience of investment projects (Dixit & Pindyck, 1994; de Neufville & Scholtes, 2011).

The field of real options has evolved significantly over the past decades, giving rise to a diverse array of modeling approaches and types of flexibility to value, Figure 2. The RO field initially appropriated the Black–Scholes formula directly from financial option theory (Black & Scholes, 1973) so that an entire project was conceptualized as a single option on one source of uncertainty – its own future value. This foundational approach treated the "investment" as a now-or-never decision under simplified assumptions. Since then, the field has expanded its degrees of freedom to include other options, more uncertainty sources, and more flexible valuation techniques (Trigeorgis, 1996; Wang & de Neufville, 2005; Martínez Ceseña et al., 2013).

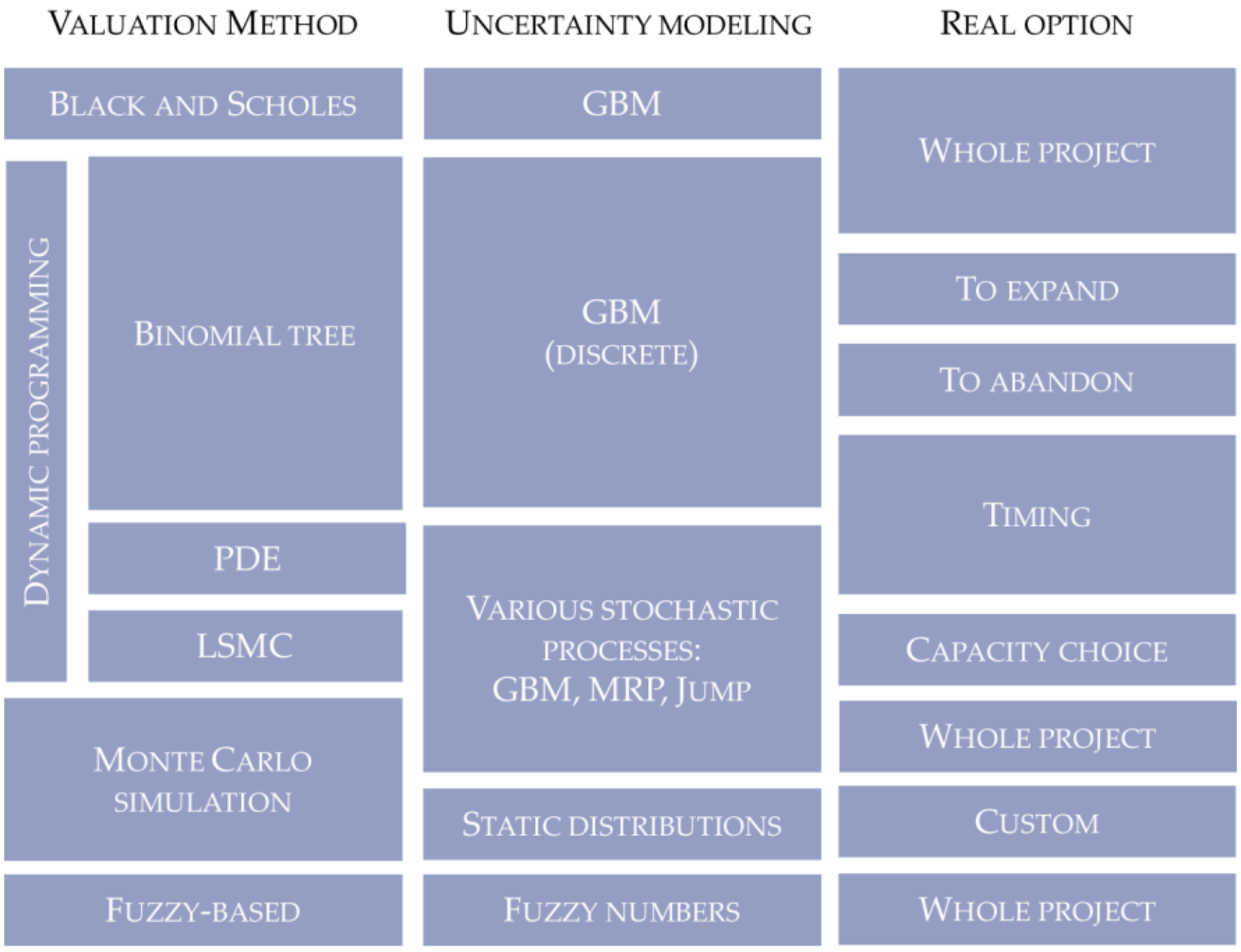

**Figure 2.** Modeling approaches in real options literature, organized by valuation technique, uncertainty type, and option structure. Rows represent common configurations.

The modern real options field is currently dominated by dynamic programming frameworks (Trigeorgis & Tsekrekos, 2018) which involve backward induction to determine the optimal decision policy under uncertainty. Dynamic programming underpins several key methods for solving real options problems, including binomial trees, partial differential equations (PDE), and least-squares Monte Carlo (LSMC). Binomial trees discretize both time and uncertainty by typically assuming a geometric Brownian motion (GBM). They are well-suited for valuing options to delay and are sometimes complimented with options to abandon and to expand (Cox, Ross, &

Rubinstein, 1979, Rambaud & Perez, 2017). PDE-based approaches are used for continuous-time problems and allow closed-form or numerical solutions under specific conditions – especially when dealing with a single uncertainty source (Dixit & Pindyck, 1994). LSMC (Longstaff & Schwartz, 2001) combine Monte Carlo simulation with regression to estimate continuation values. This enables the analysis of a wider range of options under multiple sources of uncertainty. While these techniques differ in their computational structure and flexibility, they all share the dynamic programming logic of recursively solving for optimal strategies over time.

Several additional techniques have also been adapted for real option valuation. Traditional Monte Carlo simulation can be used to estimate the value of generic or custom-designed options by comparing simulation outcomes with and without the embedded flexibility (Klepáč et al., 2013, Deng et al., 2013), or by treating the entire project as an option (Mathews, 2009, Kozlova et al., 2016). Fuzzy set theory has been employed as a non-probabilistic alternative that replaces random variables with fuzzy numbers that can incorporate the numeric assumptions of managers while simultaneously reducing computational demands (Carlsson & Fullér, 2001, Dehghani et al., 2014, Stoklasa et al., 2024). In more specialized contexts, custom optimization formulations (including mixed-integer and stochastic programming) are used to model complex project-specific flexibilities that may not be easily captured under standard option frameworks (de Neufville & Scholtes, 2011, Martínez Ceseña et al., 2013).

## 4. Sensitivity analysis: from A to Z

For the majority of non-SA researchers and practitioners, the term 'sensitivity analysis' is limited to a simple exercise of changing one variable at a time while the others remain fixed (Saltelli et al., 2019, Kozlova, Lo Piano & Yeomans, 2024). However, there is an entire scientific field of 'global sensitivity analysis' that is not normally included in the teaching of simulation and design-of-experiment which remains effectively undiscovered by those operating in applied modeling-related disciplines (Tian, 2013, Lo Piano & Benini, 2022, Puy et al., 2025). The landscape of sensitivity analysis methods (both local and global) is depicted in Figure 3 and discussed method-by-method in this section.

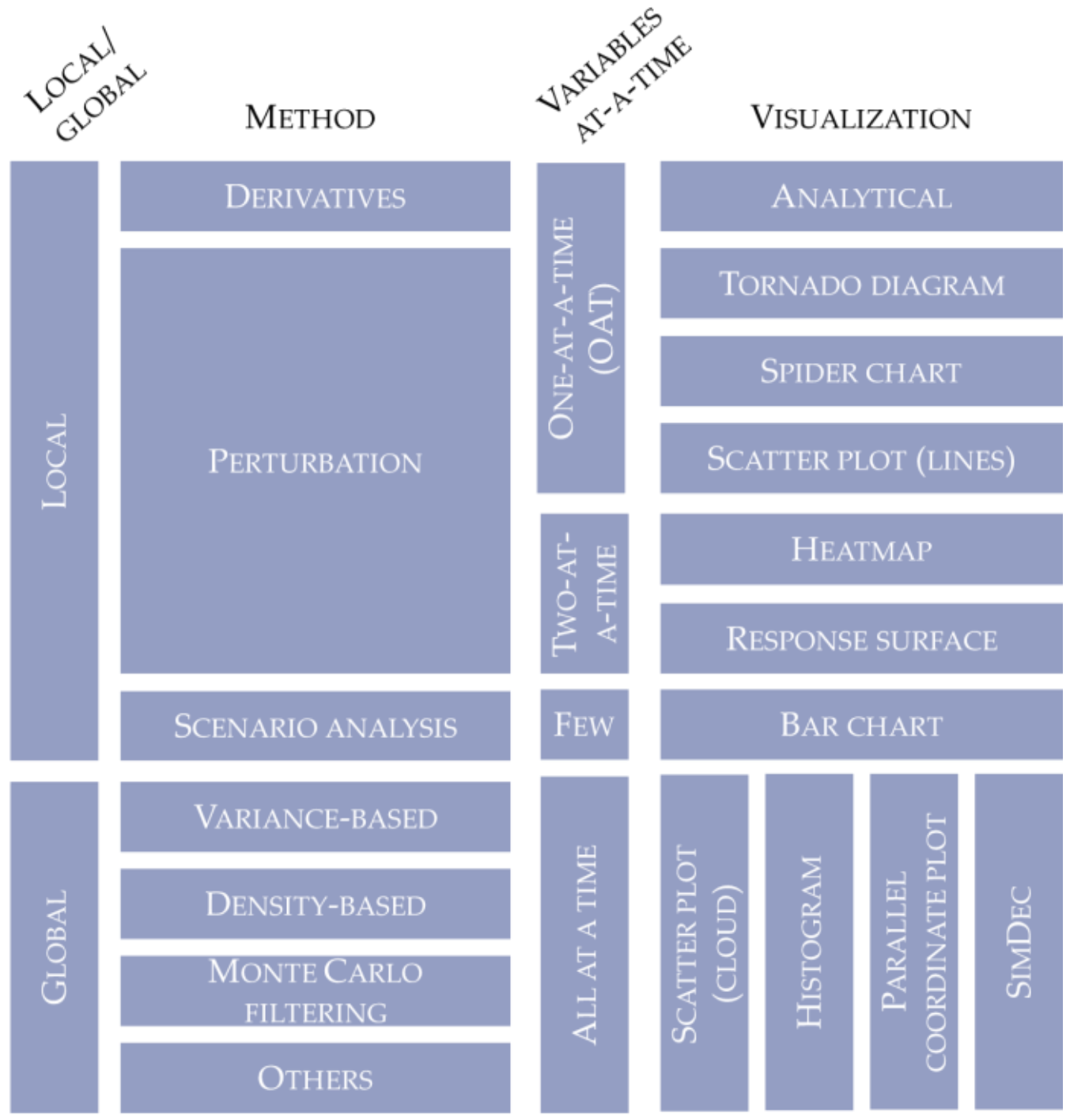

**Figure 3**. Typology of sensitivity analysis approaches and commonly used with them supportive visualizations.

### 4.1. One-at-a-time (OAT) sensitivity analysis

The simplest and most common way to conduct sensitivity analysis is by changing one-factor-at-a-time and registering the respective change in the output. The results can then be presented either in tabular format (Liu et al., 2019, Agaton, 2022, Brandon et al., 2024) or using some other characteristic visualization method, Figure 4.

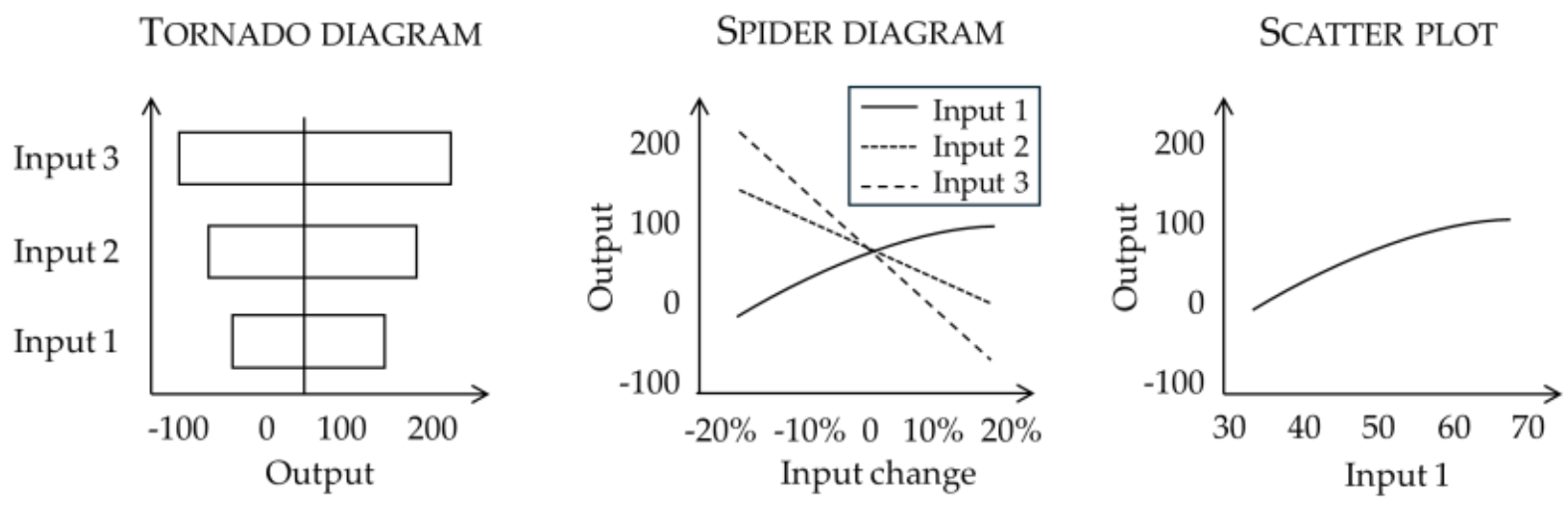

**Figure 4**. Schematic graphs for one-at-a-time sensitivity analysis.

Tornado and spider diagrams are two classic charts specifically created for OAT SA. Tornado diagrams require two valuation points for each input variable subjected to SA, and the resulting deviation of the output is illustrated with bars (Bigestans et al., 2023, Rosenberg et al., 2024, Lee et al., 2021). One limitation of tornado diagrams is that they cannot detect nonlinear relationships. Spider diagrams show the output as a function of percentage changes to the values of input variables (Woo et al., 2024, Bullock et al., 2021, Huang et al., 2020). Spider diagrams can circumvent the shortcoming of tornado diagrams by using more than two points per input to show nonlinear behaviors. However, their use introduces another limitation: all input variables need to deviate at the same percentage levels, which rarely reflects realistic situations. Regular scatter plots are frequently used to visualize OAT SA, as well (Xu et al., 2024, Thoma, 2023, Zhao et al., 2023). Unlike the cloud of dots from a Monte Carlo simulation, the resulting scatter plot is a line because it is an OAT relationship. The major problem with scatter plots in OAT SA is that, unless the inputs have similar units and numeric ranges, separate graphs have to be used to illustrate each input.

OAT SA can also be performed analytically by computing partial derivatives of the output with respect to individual input variables (Tolga, 2020, Yin & Hu, 2007, Lee et al., 2008, Wu, 2011). This approach provides a local measure of sensitivity around a nominal point and is often used in models with closed-form solutions or symbolic representations. It does not require visualization for interpretation, although graphical tools may still aid in its communication.

The crucial limitation of all OAT approaches is that they fail to account for any joint effects between the variables. Even in simple formulations such as the ubiquitous *quantity* × *price* structure in real options models, the interaction between variables leads to fundamentally different outcomes compared to additive relationships. Since OAT methods cannot capture even the most basic nonlinearities, they completely overlook all of the more complex interaction effects that arise in real-world problems (Saltelli & Annoni, 2010, Pianosi, et al., 2016).

### 4.2. Two-at-a-time (TAT) sensitivity analysis

Two-at-a-time sensitivity analysis is performed by perturbing the input values as in OAT, but changing two inputs simultaneously instead of just one. Namely, for each value of one input, every value of the other input is varied and the corresponding output value is recorded. This results in a 2D array of the output values.

The results from TAT SA can be reported in tabular format (Di, 2020, Lin et al., 2019, Reindorp & Fu, 2011), however, the interpretability deteriorates with more points for each input. Instead, classic TAT visualizations such as heatmaps (Zetterholm et al., 2022, Truong et al., 2024, Guo & Zhang, 2020) and response surfaces (Kim et al., 2017, Huang et al., 2014, Chen, 2012, Park & Lim, 2009, Çagri et al., 2010) can be applied to improve the readability of TAT results, Figure 5.

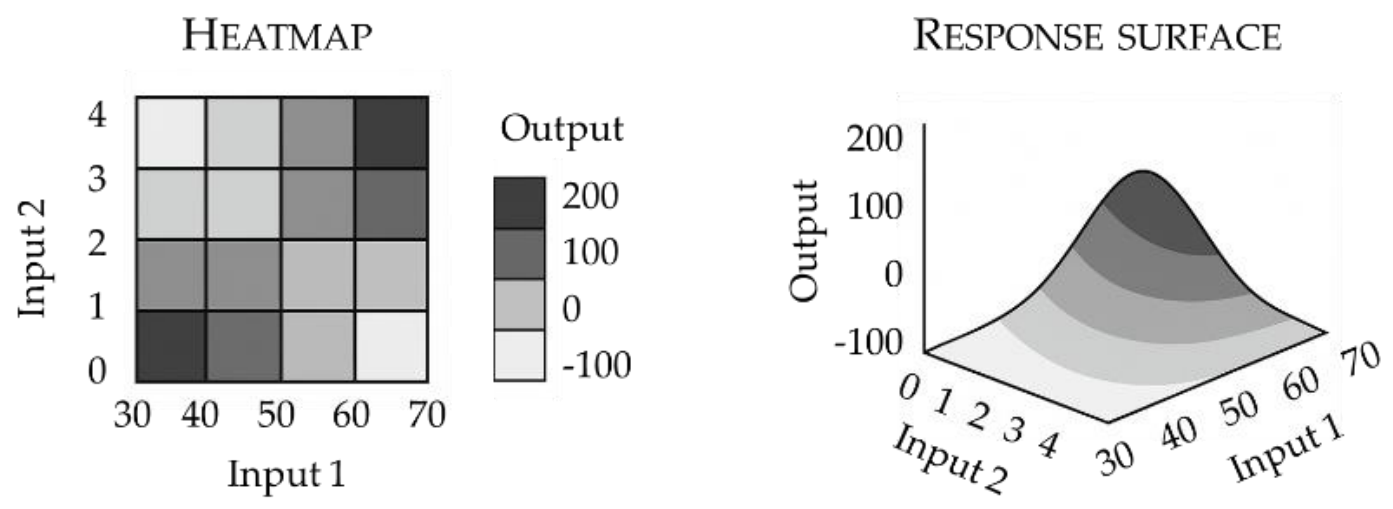

**Figure 5**. Schematic graphs for two-at-a-time sensitivity analysis.

Heatmaps employ one input value on the X-axis, another one on the Y-axis, and the corresponding output values are represented as color gradients on the mesh between the axes. Response surfaces introduce a third Z-axis for the output and, thereby, extend the visualization into 3D space.

While TAT SA are able to capture interactions between pairs of input variables, achieving this comprehensively requires repeating the analysis for every pair – which becomes increasingly impractical as the number of inputs grows. Moreover, this approach still leaves much of the output space unexplored, particularly when more than two variables are changing simultaneously. As a result, TAT analysis can only be classified as a local SA method and is generally considered insufficient for a thorough, global exploration of model behavior (Saltelli et al., 2008, Iooss & Lemaître, 2015).

### 4.3. Scenario analysis

One way to change more that two variables at a time is scenario analysis. A custom set of a few scenarios in which a few input variables are changed produces several output values which can then be visualized in a bar chart (Banda, 2020, Delbridge & King, 2016). One drawback to scenario analysis is that it remains unclear which specific input variable led to the actual variation in the output. Another limitation is that, because only a restricted number of scenarios are checked, the method remains rooted within the group of local approaches.

### 4.4. Global sensitivity analysis

Global methods are based on Monte Carlo simulation (or other types of sampling) in which the overarching goal is to explore the entire space of possible outputs by varying all of the inputs simultaneously. Data collected from the simulation is used to compute importance measures.

Instances in which a Monte Carlo simulation is conducted without any follow-up analysis of its input–output data (Kemala & Hakam, 2025, Sim, 2023, Freitag et al., 2020, Lee et al., 2021) do not strictly qualify as global sensitivity analysis. Nevertheless, using Monte Carlo simulation to complement one-at-a-time (OAT) analysis is considered preferable to relying on OAT alone.

While many types of GSA exist and continue to be developed (Tarantola et al., 2024), we will subsequently describe only the variance-based, moment-independent, and Monte Carlo filtering groups in any detail. For a more comprehensive introduction to the field of GSA, interested readers are directed to Saltelli et al. (2004).

#### *4.4.1. Sampling strategies*

In GSA, the process of generating input data is referred to as its sampling strategy. Classic Monte Carlo simulation represents the baseline approach and is referred to as simple random sampling. However, simple random sampling does not ensure an even coverage of the input space, which can result in lower precision when estimating importance measures (Saltelli et al., 2008, Iooss & Lemaître, 2015). More structured approaches have been developed to improve sampling efficiency. One widely recommended method is quasi-random sampling (or quasi-Monte Carlo) where each new point is selected based on the location of previously sampled points. This approach uses low-discrepancy sequences (such as Sobol' or Halton) to achieve a more even distribution throughout the input domain (Owen, 2019). Another option is the full factorial design which is a deterministic technique where all combinations of input levels are evaluated on a multi-dimensional grid. While full factorial designs guarantee complete coverage, they become computationally intractable for higher numbers of inputs and evaluation points due to the curse of dimensionality (Montgomery, 2017).

#### *4.4.2. Variance-based GSA*

Various methods exist for estimating the importance of input variables based on input-output data. Variance-based sensitivity analysis (Sobol' indices, in particular) is the most common and only approach used in the RO literature (Gkousis et al., 2024, Guo & Zhang, 2020, Araya et al., 2021, Liukkonen et al., 2025). Sobol' indices show what portion of the output variance can be attributed to a particular input variable or some combination of inputs (Sobol, 1993). First-order indices show individual contributions of each input variable, while total indices capture all the interactions the input variable has with the other variables. The most widely-used numeric estimators for Sobol' indices were introduced in Saltelli (2002) and Saltelli et al. (2010). However, these estimators require a specific design of experiments and they cannot process dependent data. Consequently, various authors have introduced "workarounds" that enable computing Sobol' indices directly from the input-output data (Marzban & Lahmer, 2016, Kucherenko & Song, 2017, Kozlova et al., 2025).

#### 4.4.3. Moment-independent GSA

Moment-independent GSA evaluates how much the output distribution changes when a given input is fixed, without relying on variance-based or moment-based assumptions (Borgonovo, 2007, Da Veiga, 2015, Puy et al., 2024). These methods are better suited for complex, nonlinear models with nonstandard output distributions (e.g., multi-modal, skewed, or heavy-tailed). However, the resulting measures are expressed as a statistical distance or divergence, which makes them less intuitive to interpret compared to variance-based sensitivity indices.

#### 4.4.4. Monte-Carlo filtering

Monte Carlo filtering is another GSA method that identifies influential input variables by comparing their distributions across different output-based subsets. After running a Monte Carlo simulation, outputs are classified into behavioral and non-behavioral groups (e.g. success vs. failure), and the input distributions for each group are compared using statistical tests like the Kolmogorov–Smirnov test. Inputs whose distributions differ significantly between the groups are considered influential. This method is especially useful when the model output is binary, threshold-based, or hard to reduce to a scalar summary (Saltelli et al., 2008).

#### 4.4.5. Visualizations for GSA

No standard visualization practices exist for GSA. GSA is often used either without supporting visualization or with only a bar chart or a box plot showing the computed sensitivity indices, rather than a visualization of the shape of the underlying relationship (Kozlova et al., 2025). A standard visualisation for Monte Carlo simulation is a histogram showing the distribution of the output values (Freitag et al., 2020, Kemala & Hakam, 2025). However, these figures also lack an explicit link to the inputs, Figure 6.

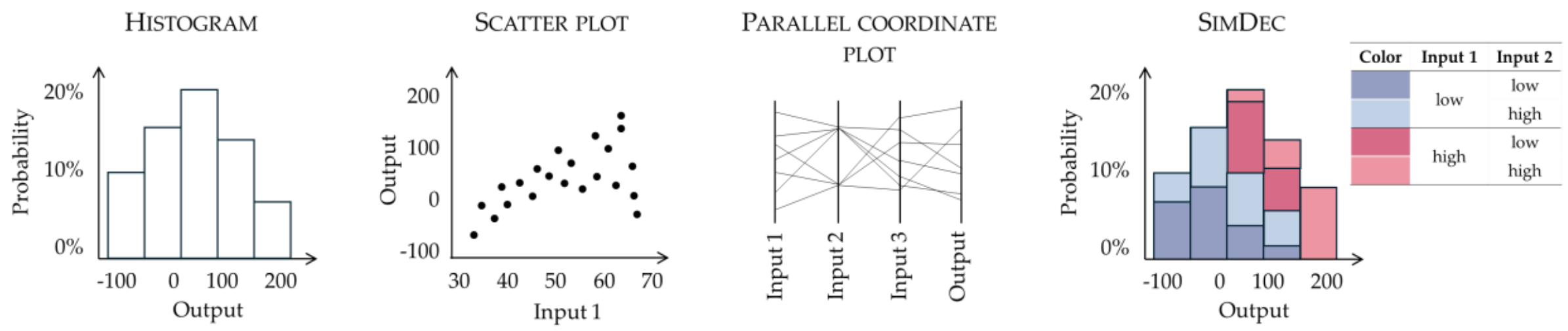

**Figure 6**. Schematic types of visualizations used in support of global sensitivity analysis.

A classic scatterplot could be used to visualize the relationships between inputs and outputs (Guo & Zhang, 2020). However, scatterplots are limited to showing one input per graph, require many graphs to explore all input-output linkages, and miss all potential second-order heterogeneous effects (Kozlova et al., 2024).

Parallel coordinate plots resolve the dimensionality limitation by aligning the axes for each input with the output using parallel vertical coordinates. In the plots, each multidimensional data point is represented by a line connecting all of the coordinates together (Pianosi et al., 2015). However, this type of visualization features an overcrowding from the thousands of GSA samples (Heinrich & Weiskopf, 2013). Furthermore, the order of plotting may confound influential connections. This ordering hindrance is illustrated in Figure 6, where the perception of the prominent influence from Input 2 on the Output is obstructed by the positioning of Input 3 between them.

The SimDec method provides another approach for visualizing multidimensional input-output relationships (Kozlova et al., 2024). SimDec operates in conjunction with the computation of sensitivity indices and decomposes the simulation data using combinations of the states of the most influential inputs. The resulting subgroups are mapped onto the output distribution by color-coding series of a stacked histogram. SimDec also contains a boxplot representation to visualize skewed or otherwise poorly readable distributions. Furthermore, SimDec possesses a two-output plot for problems in which two outputs matter. These plots show not only a decomposition of the input-output relationships, but also how the two outputs can be interlinked via a scatter subplot. The SimDec procedure has been made universally available via open-source packages in several languages (Kozlova et al., 2024) together with a no-code web dashboard (Roy & Kozlova, 2024).

## 5. Review of literature

### 5.1. Review methodology

The search of publications was conducted using the SCOPUS database in June 2025. The search possessed the following criteria: (i) *real options* literature that mention (ii) *sensitivity analysis*, limited to (iii) *journal* articles in *English* only. The following query has been used:

```
( TITLE-ABS-KEY ( "real option*" ) AND TITLE-ABS-KEY ( "sensitivity analysis" ) )
AND ( LIMIT-TO ( DOCTYPE , "ar" ) ) AND ( LIMIT-TO ( LANGUAGE , "English" ) )
```

The search returned 193 documents. Each document was analyzed in accordance with the attributes described below. Out of the initial selection of articles, 49 were discarded due to no access to the full text (not Elsevier, not open access, and not shared via ResearchGate). Thirteen more were discarded as ‘out of scope’, as they were either theoretical papers, or surveys of corporate practices, or literature reviews. The resulting sample consisted of 131 paper, Figure 7.

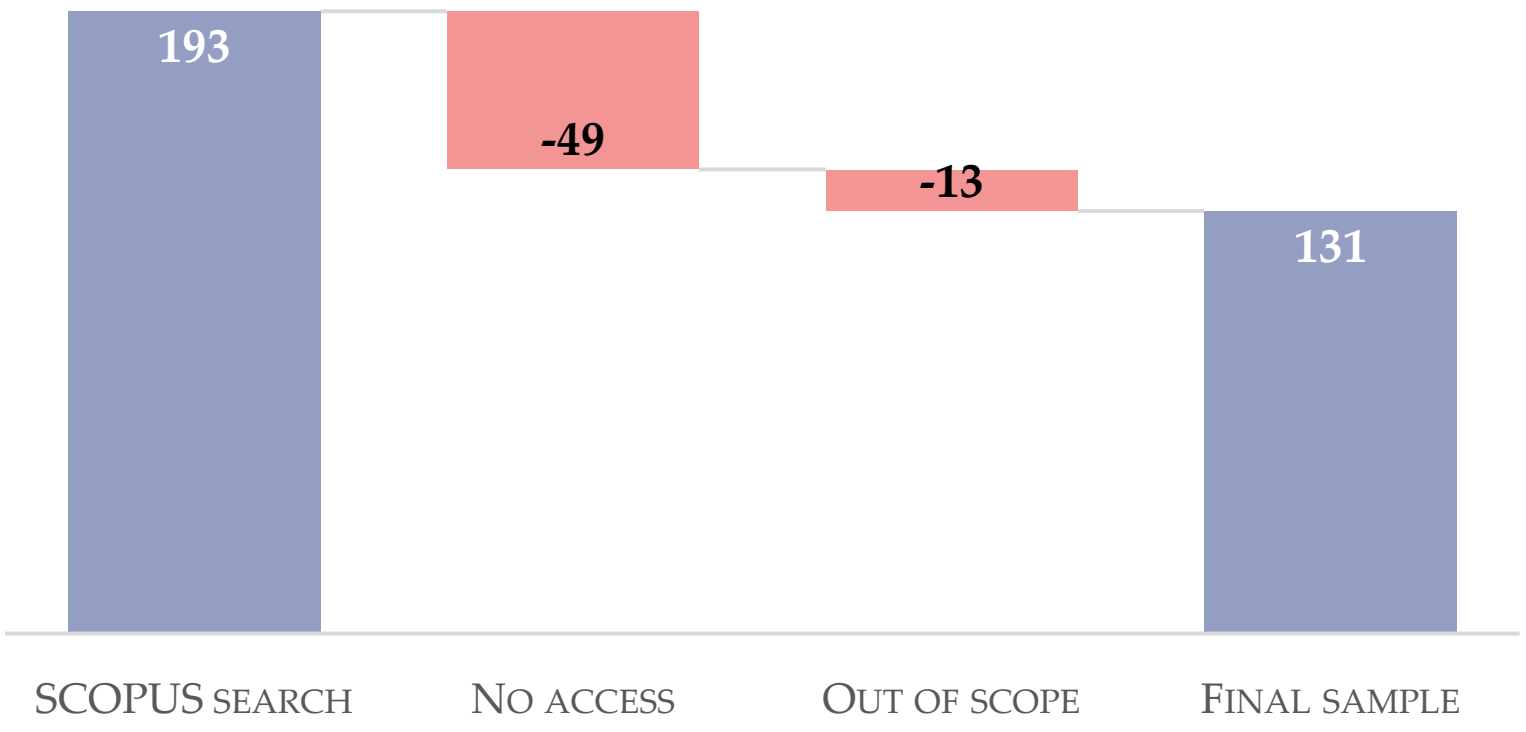

**Figure 7.** Paper selection flow.

The sample was analyzed along the two major dimensions: (i) *RO study design* and (ii) *sensitivity analysis*, across the following attributes:

- *RO study design*: valuation method; uncertainty sources; uncertainty modelling; number of uncertainty sources; real option type; number of real options.
- *Sensitivity analysis*: type of analysis; number of methods used; inputs to SA; types of inputs to SA; number of inputs to SA; outputs for SA; whether scenarios are considered; visualisation type.

Each attribute has several categories which were assigned for each paper. The exhaustive list of categories with short explanations is provided in Appendix A. The resulting dataset of reviewed papers is supplied in Appendix B.

### 5.2. Application areas

The selection of papers showcases a diverse range of application areas, Figure 8, that demonstrate the ubiquitous applicability and nature of sensitivity analysis usage. The leading category, 'energy', is also represented by a variety of subtopics, including electricity generation, distribution, storage, consumption, and others.

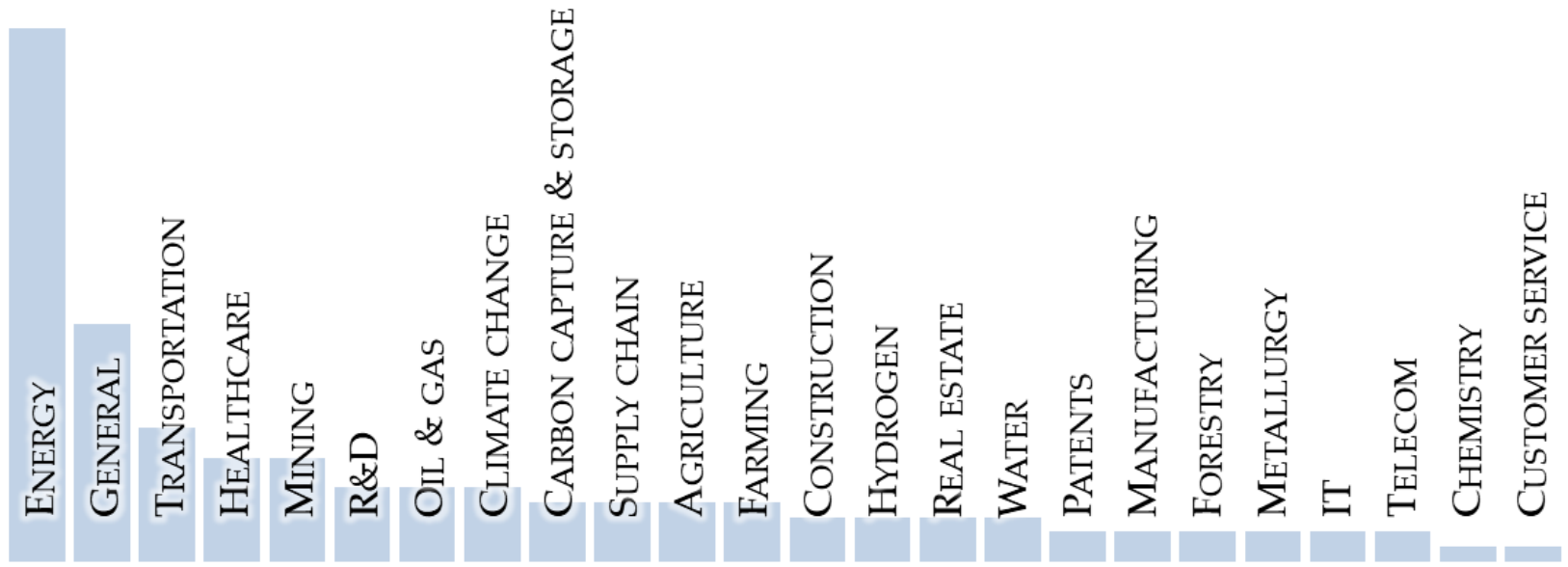

**Figure 8.** Application areas of the selected papers.

This thematic distribution approach can be directly compared with the only theme-agnostic review of real options literature by Trigeorgis & Tsekrekos (2018). In their review, the energy category (that includes the topics of mining, land use, and climate change shown separately here) ranks fifth place after the Investment, R&D, Manufacturing, and Supply chain themes. This contrast indicates that researchers working with energy applications are more accustomed to sensitivity analysis than others.

### 5.3. Sensitivity analysis practices

An examination of the sensitivity analysis dimension reveals the prevalence of the following practice. The OAT approach is used as a solo method without considering different scenarios and to illustrate results in a series of scatter plots, where the sensitivity of 1-2 outputs are checked with 2-3 model parameters (e.g., Rios et al., 2019, Adkins & Paxson, 2017, Azimi et al., 2013, Tseng & Lin, 2007), Figure 9.

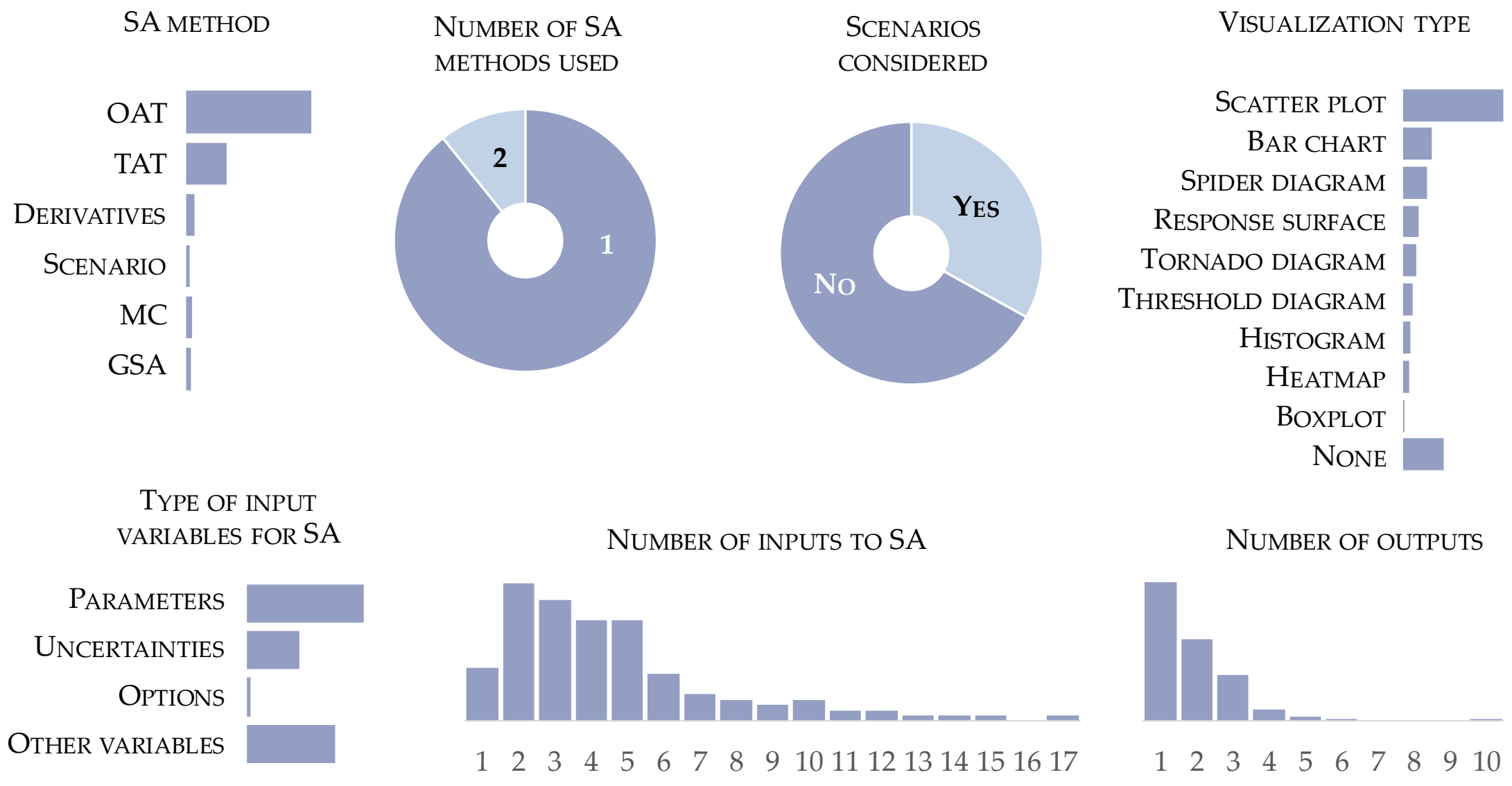

**Figure 9**. Attributes of the sensitivity analysis dimension across reviewed papers. Horizontal bar charts represent non-exclusive attributes (multiple categories per paper), while histograms and doughnut charts are used for exclusive attributes (single category per paper).

The OAT approach dominates the field with 66% of papers employing it, followed by TAT at 22%. All other approaches are chosen in only a small minority of the papers. As already noted in Introduction, GSA is used in only 4 papers (Gkousis et al., 2024, Guo & Zhang, 2020, Araya et al., 2021, Liukkonen et al., 2025). In 89% of the papers, only one type of sensitivity analysis is used, while no paper employs more than two methods. One third of the papers repeat their analysis approach over several scenarios or cases.

The majority of the papers (81%) portray their sensitivity analysis results in visual format. Scatter plots represent the most common vehicle to display OAT results (40%), while classic spider charts and tornado diagrams are employed 20% and 16% of the time, respectively. Bar charts are frequently used (20%) for OAT when only a few points of an input (e.g. four levels of volatility) are checked. Heatmaps and response surfaces are exclusively used for TAT. A specific, customized threshold diagram showcasing decision regions is used in several studies and repeated over several perturbations of one- or two- variables at-a-time (Abdul-Salam et al., 2022, Mei & Wetzstein, 2017, Hampe & Madlener, 2016, Chladná, 2018, Delbridge & King, 2016, Chladná, 2007).

In terms of the inputs to sensitivity analysis, the most frequent choice in 67% of the papers is to test model parameters (e.g. drift and volatility in GBM processes). The second most frequent choice (51% of the time) is to consider other variables that have been fixed in the model, such as investment costs or different subsidy levels. Sometimes (31% of the time) the uncertainties, themselves, are subjected to a sensitivity study (e.g. the output values are checked against a range of values for uncertain price and demand). Only three papers in the sample examined the sensitivity (namely, the scale of expansion) to the real option itself (Durica et al., 2018, Li & Cao, 2022, Guo et al., 2024). Two thirds (66%) of the papers analyzed from 2 to 5 inputs, while only 6% of them exceeded 10 inputs. Although a single-studied output designs represent the most popular arrangement (48%), two-output studies (28%) were not uncommon (the most prevalent being option value and optimal timing outputs within timing real options). Only 7% of papers considered more than 3 outputs, which stemmed from either having several real options (Maier, 2021, Hampe & Madlener, 2016, Guthrie, 2012), comparing different approaches (e.g. NPV alongside option value) (Zhang et al., 2018, Hampe & Madlener, 2016), or involving some other specific problem formulations (Ribeiro et al., 2020, Zhang et al., 2019, Rau & Spinler, 2017).

Overall, it was observed that the RO literature is skewed heavily (94% of the sample) toward local approaches for sensitivity analysis. However, since the details of each analysis design vary greatly, no "gold standard" has been shown to exist for the RO field.

### 5.4. Linking sensitivity analysis to RO modeling choices

It was also interesting to determine if the choices underlying the specific RO study design actually influenced the methodological arrangement of its sensitivity analysis. Although no rigorous analysis could be performed due to the small amount of data in several of the categories (see

Figure 9, SA method), the visual composition of the methodological choices sheds light on their co-design, Figure 10.

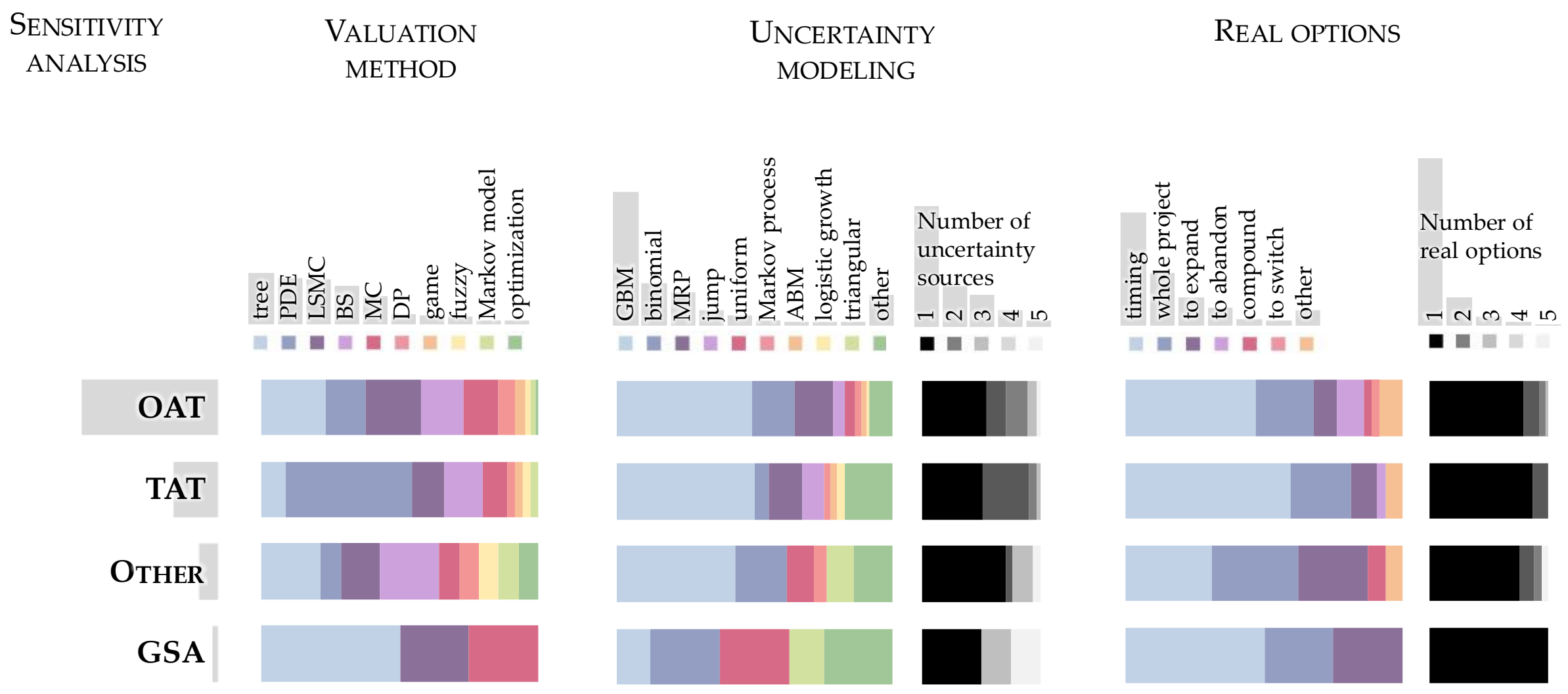

**Figure 10**. RO study design choices across different types of sensitivity analysis in a 100% stacked bar representation. Total frequences of the study design element categories are depicted with gray column charts with aligned Y-axes, so that the heights of the columns across the graphs are comparable.

All SA types contain a variety of approaches based upon valuation method. TAT is applied more frequently in PDE studies than other SA types, but this is more likely to be a coincidence rather than a methodological tendency. Notably, 'Other' SA types (Derivatives, Scenario, and Monte Carlo simulation) are used more than others in conjunction with non-mainstream RO approaches such as fuzzy-based methods, Markov switching models, and non-DP optimization. GSA, despite its methodological complementarity with Monte Carlo simulation, has been selected for other study designs too, including binomial trees and LSMC.

With respect to uncertainty modelling, the movement from OAT to GSA has been characterized by the diminishing prevalence of GBM and by a more pronounced share of other types of uncertainty modelling. Although classic GSA is performed with static distributions, the sample indicates that GSA has been successfully employed for dynamic uncertainties modelled by stochastic processes. In terms of the number of uncertainty sources, GSA has the lowest share of single uncertainty source studies. To some extent, this observation can be explained by the ability of the method to capture multiple varying inputs simultaneously.

The two most commonly used real options (timing and whole project) dominate all types of sensitivity analysis. Thus, the number of real options used does not display any particular pattern across sensitivity analysis types and is heavily dominated by one option per study.

Overall, no pronounced patterns appear at the cross-section of RO and SA study designs, except that both independently gravitate towards simpler alternatives. This results in (i) more OAT than GSA in sensitivity analysis, and (ii) a lower number of uncertainties and generic options considered in main study designs.

### 5.5. Discussion and recommendations

In summary, RO literature have relied heavily on local approaches of sensitivity analysis – particularly on OAT methods. This myopic reliance prevents a more thorough model investigation because it ignores the joint effects of input variables and their possible interactions. This limitation has been widely discussed throughout the sensitivity analysis literature (Saltelli et al., 2019, Borgonovo & Plischke, 2016, Saltelli et al., 2008; Iooss & Lemaître, 2015) and its impacts have been increasingly recognized in applied modeling fields such as finance and economics (Pianosi et al., 2016; Baroni & Tarantola, 2014). In the context of RO studies, this reliance on local methods can obscure the understanding of complex, nonlinear relationships that influence option value and decision robustness (de Neufville & Scholtes, 2011; Wang & de Neufville, 2005). The situation is worsened further when only a few input variables are subjected to sensitivity analysis, thereby leaving the overall model behavior largely uninvestigated.

Global sensitivity analysis has been increasingly recommended to thoroughly explore model behavior in complex decision models (Saltelli et al., 2004, Iooss & Lemaître, 2015, Pianosi et al., 2016). While RO models often involve stochastic processes and dynamic system behaviors, GSA typically relies on Monte Carlo simulation and assumes input variables follow static probability distributions. However, this does not mean that GSA is incompatible with RO analysis. In practice, Monte Carlo simulation can be conducted on top of an RO model by treating the parameters of the underlying stochastic processes and other variables as uncertain inputs represented by static distributions. This layered approach enables the generation of input–output datasets suitable for GSA, while preserving the dynamic nature of the RO framework. A handful of RO studies have already demonstrated this workflow effectively (Gkousis et al., 2024, Guo & Zhang, 2020, Araya et al., 2021), and one study explicitly illustrates the complementarity of methods by reusing Monte Carlo data both for real option valuation and for global sensitivity analysis (Liukkonen et al., 2025). Incorporating GSA into RO modeling not only improves transparency in how uncertainties and assumptions influence valuation outcomes, but also enables the identification of key drivers supporting more informed decision-making under uncertainty.

By itself, however, GSA is limited to quantitative procedures for computing importance measures. While identifying the strength of the input–output relationships is informative, understanding their shape can be equally crucial. This is particularly true in decision-making contexts where nonlinearities, thresholds, and interaction effects may lead to qualitatively different outcomes (Kozlova et al., 2024, Saltelli & Annoni, 2010, Pappenberger et al., 2008). Scatter plots provide a convenient means for exploring bivariate relationships, but require multiple concurrent plots to capture all of the influential input variables. Multidimensional visualization techniques such as parallel coordinate plots (Wegman, 1990, Pianosi et al., 2015) and SimDec (Kozlova & Yeomans, 2024) enable interpretable inspections of multi-input relationships and nested interactions. Incorporating visual diagnostics alongside numerical GSA outputs significantly enhances the informativeness of decision support.

To enable easy access and use, researchers have been developing open-source GSA libraries and software packages programmed in multiple computer languages. These include Python library SALib (Herman & Usher, 2017), sensitivity analysis packages in R (Iooss & Lemaître, 2015, Da Veiga et al., 2021), SAFE toolbox for Matlab (Pianosi et al., 2015), GSA package in Julia (Dixit & Rackauckas, 2022), and an open-source project for SimDec in Python, R, Julia, and Matlab (Kozlova et al., 2024).

Beyond the practice and implementation of sensitivity analysis, one underdeveloped direction in RO research concerns the actual modeling foundations, themselves. The majority of reviewed studies continue to rely on a single, generic real option – the option to defer an investment. This focus contrasts with the original philosophical foundation of RO theory, which sought to move beyond deterministic evaluation by capturing the value of flexibility under uncertainty (Trigeorgis, 1996; Dixit & Pindyck, 1994). Several papers have advocated for a transition toward project- or technology-specific flexibilities, which has often been described as “real options in-project” (de Neufville & Scholtes, 2011; Wang & de Neufville, 2005, Ceseña et al., 2013). This shift is in line with a clarion call to rethink overly-constrained models in economics (DeMartino et al., 2024). Following this pathway, future research should involve building models containing multiple potential flexibilities. These flexibilities can be randomly activated and deactivated during Monte Carlo simulation. A subsequent GSA will assess which flexibilities, or their combinations, would be most valuable under different uncertainty scenarios which would help to guide the design of robust and adaptive strategic pathways. For models in which the behavior is highly nonlinear and interdependent, the use of GSA jointly with multidimensional visualization techniques would become especially important in order to reveal the complex interaction effects and to identify key leverage points.

## 6. Conclusion

This study has systematically reviewed the use of sensitivity analysis in real options literature to assess the prevalence, quality, and type of methods employed. The goal was to identify whether the RO field leverages global sensitivity analysis techniques that allow for a more comprehensive understanding of model behavior under uncertainty. One hundred and thirty-one peer-reviewed journal articles were examined, and their SA practices, modeling approaches, and application areas were categorized.

This review confirmed that SA remains underdeveloped throughout all RO literature, with most studies relying either on simplified one-at-a-time approaches or omitting SA altogether. The RO community is therefore encouraged to make global sensitivity analysis the standard component of model evaluation, in order to strengthen the overall methodological rigor of the field. RO research is already grounded in advanced mathematical and computational techniques, which suggests that the adoption of more rigorous global sensitivity analysis methods should not pose a major impediment or technical barrier. The ready availability of open-source packages further facilitates the widespread adoption of GSA. Integrating GSA into standard modeling practice would significantly enhance the overall transparency, credibility, and practical relevance of RO-based analyses.

## Acknowledgments

The authors are grateful for the support of OP Foundation, grant number 20250110.

## Appendix A. Review attributes and categories.

| Review attribute | Categories |
|---|---|
| **Application area** | agriculture, CCS (carbon capture and storage), chemistry, climate change, construction, customer service, energy, farming, forestry, general, hydrogen, IT (information technology), manufacturing, medical, metallurgy, mining, oil&gas, patents, R&D, real estate, supply chain, telecom, transportation, water. |
| **Real options study design** | |
| Valuation method | BS (Black and Scholes), PDE (partial differential equations), LSMC (least-squared Monte Carlo), DP (dynamic programming), tree (binomial/decision), game (game theoretic approach), Markov model, MC (Monte Carlo simulation), optimization, fuzzy (fuzzy set theory-based approaches), other. |
| Uncertainty modelling | GBM (geometric Brownian motion), MRP (mean reverting process), jump (jump process), Markov (process/chain), BL (binomial lattice), triangular, normal, fuzzy, lognormal, uniform, trapezoidal, and others. |
| Number of uncertainty sources | numeric |
| Real option type | timing (optimal timing, to delay, to defer), whole project (to invest), to expand, to abandon, to pause, to stop, capacity choice, to switch, other. |
| Number of real options | numeric |
| **Sensitivity analysis configuration** | |
| SA method | derivatives, OAT (one-at-a-time perturbations), TAT (two-at-a-time perturbations), scenario, MC (Monte Carlo simulation), GSA (global sensitivity analysis). |
| Number of methods used | numeric |
| Type of variables for SA | P (parameters), U (uncertainties), O (options), OV (other variables). |
| Number of inputs to SA | numeric |
| Number of outputs for SA | numeric |
| Scenarios/cases | yes, no |
| Visualization for SA | no, scatter plot, bar chart, spider (spider diagram), tornado (tornado diagram), heatmap, response surface, histogram, threshold diagram, other. |

## Appendix B. Collected review data.

| Paper | Application area | Valuation method | Uncertainty modelling | Number of uncertainty sources | Real option type | Number of real options | SA method | Number of methods | Type of variables for SA | Number of inputs to SA | Number of outputs for SA | Scenarios/cases | Visualization for SA |
|---|---|---|---|---|---|---|---|---|---|---|---|---|---|
| (Guo et al., 2024) | transportation | LSMC | GBM | 1 | to expand, timing | 2 | derivatives | 2 | O, P | 8 | 2 | no | bar chart, scatter plot |
| (Dib et al., 2025) | energy | LSMC, tree, other | GBM, jump | 1 | timing, to invest, other (demand response (de)activating) | 3 | OAT | 1 | OV | 1 | 2 | yes, scenarios | scatter plot |
| (Triki & Abid, 2025) | general | PDE | jump, Markov | 2 | timing | 1 | TAT | 1 | OV, P | 6 | 3 | yes, financing policies, regimes | scatter plot |
| (Song et al., 2023) | R&D | game, PDE | GBM, other | 3 | timing, other (to cooperate) | 2 | OAT, TAT | 1 | P, OV | 9 | 1 | no | spider, response surface |
| (Liu & Zhu, 2023) | construction | BS | GBM | 1 | whole project | 1 | OAT | 1 | P | 4 | 3 | no | scatter plot |
| (Ho et al., 2023) | general | PDE | not specified | 1 | whole project | 1 | OAT | 1 | OV | 1 | 1 | no | scatter plot |
| (Cao et al., 2025) | general | PDE, LSMC | MRP | 1 | timing | 1 | TAT | 1 | P | 3 | 2 | no | scatter plot |
| (Kemala & Hakam, 2025) | oil&gas | tree | BL | 1 | abandon, contract, expand | 3 | OAT, MC | 2 | P, OV | 13 | 1 | yes, scenarios | tornado, histogram |
| (Lee et al., 2022) | medical | Markov model | Markov | 4 | option to access future treatment | 1 | OAT, TAT | 2 | P, OV | 10 | 1 | no | tornado, heatmap |
| (Li et al., 2023) | climate change | BS, tree | empirical, BL | 3 | timing | 1 | OAT | 1 | P | 2 | 3 | yes, scenarios, projects | scatter plot |
| (Zetterholm et al., 2022) | energy | MC | GBM, jump, other | 4 | timing | 1 | MC | 1 | U | 4 | 4 | yes, technology | heatmap, boxplot |
| (Truong et al., 2024) | climate change | tree | extreme value, ABM | 2 | timing | 1 | OAT, TAT | 2 | P, U | 5 | 2 | yes, projects | heatmap, none for OAT |
| (Sim, 2023) | energy | BS | GBM | 1 | whole project | 1 | OAT, MC | 2 | OV | 3 | 1 | yes, areas | bar chart, distribution over time from Vensim |
| (Xu et al., 2024) | metallurgy, hydrogen | LSMC | GBM | 4 | timing | 1 | OAT | 1 | P, U, OV | 6 | 2 | no | scatter plot |
| (Li et al., 2025) | climate change | optimization | custom (EnergyPlus) | 1 | timing | 1 | scenario | 1 | U | 1 | 4 | yes, scenarios | bar chart |

| Paper | Application area | Valuation method | Uncertainty modelling | Number of uncertainty sources | Real option type | Number of real options | SA method | Number of methods | Type of variables for SA | Number of inputs to SA | Number of outputs for SA | Scenarios/cases | Visualization for SA |
|---|---|---|---|---|---|---|---|---|---|---|---|---|---|
| (Tan et al., 2024) | CCS | LSMC | GBM | 1 | timing | 1 | OAT, TAT | 2 | U, OV | 5 | 3 | yes, scenarios | scatter plot, bar chart |
| (Agaton, 2022) | energy | BS | GBM, jump | 1 | timing | 1 | OAT | 1 | P, OV | 14 | 2 | no | no |
| (Thoma, 2023) | patents | tree | BL | 1 | whole project | 1 | OAT | 1 | P, OV | 5 | 1 | no | scatter plot |
| (Zhao et al., 2023) | R&D | fuzzy | GBM, trapezoidal fuzzy, jump | 3 | whole project | 1 | OAT | 1 | P, U | 5 | 1 | yes, scenarios | scatter plot |
| (Bigestans et al., 2023) | hydrogen | MC | GBM | 2 | to expand | 1 | OAT | 1 | P, OV | 5 | 1 | yes, scenarios | tornado |
| (Rosenberg et al., 2024) | medical | other | triangular | 8 | whole project | 1 | OAT | 1 | P, U, OV | 10 | 1 | yes, models | tornado |
| (Bi & Lyu, 2022) | energy | PDE | GBM | 1 | timing | 1 | OAT, TAT | 2 | P, OV | 3 | 2 | yes, subsidies | scatter plot, response surface |
| (Yang et al., 2024) | energy | LSMC | GBM | 3 | timing | 1 | OAT | 1 | U, OV | 3 | 1 | yes, areas | scatter plot, bar chart |
| (Chen et al., 2024) | energy | PDE | GBM | 4 | timing | 1 | OAT | 1 | U, OV | 4 | 2 | yes, markets | spider |
| (Fu et al., 2024) | energy | MC | GBM | 2 | timing | 1 | OAT | 1 | U, OV | 3 | 1 | yes, areas | polar plot |
| (Penizzotto et al., 2024) | energy | LSMC | GBM, jump | 3 | timing, to expand | 2 | TAT | 1 | U, P | 2 | 1 | yes, stochastic process | scatter plot |
| (Brandon et al., 2024) | chemistry | other | scenario, jump, GBM | 4 | to abandon | 2 | OAT | 1 | U, OV | 17 | 2 | yes, scenarios | no |
| (Gkousis et al., 2024) | energy | LSMC | GBM, uniform, lognormal, normal | 6 | timing | 1 | GSA (Sobol) | 1 | U | 4 | 2 | yes, scenarios | bar chart |
| (Sadat et al., 2023) | manufacturing | PDE | GBM | 1 | timing | 1 | OAT | 1 | U, P | 3 | 2 | yes, different models | scatter plot |
| (Yang et al., 2025) | hydrogen | LSMC | GBM, Markov, learning curve | 3 | timing, to expand, to stop | 3 | OAT | 1 | P | 3 | 1 | yes, investment timing | bar chart |
| (Woo et al., 2024) | transportation | tree | BL | 3 | timing, choice | 2 | OAT | 1 | P, U, OV | 6 | 1 | no | spider |
| (Pirayesh et al., 2024) | general | game, tree | regime switching, BL | 1 | capacity choice | 1 | OAT | 1 | P | 6 | 3 | yes, models | scatter plot |
| (Bullock et al., 2021) | medical | tree | PERT, uniform | 11 | to abandon | 1 | OAT | 1 | U, P | 3 | 1 | yes, technologies | spider |

| Paper | Application area | Valuation method | Uncertainty modelling | Number of uncertainty sources | Real option type | Number of real options | SA method | Number of methods | Type of variables for SA | Number of inputs to SA | Number of outputs for SA | Scenarios/cases | Visualization for SA |
|---|---|---|---|---|---|---|---|---|---|---|---|---|---|
| (Liu et al., 2019) | water | BS, MC | GBM, MRP | 1 | whole project | 1 | OAT | 1 | P | 3 | 1 | no | no |
| (Maier, 2021) | mining | PDE | GBM | 1 | to abandon, to pause | 2 | OAT, TAT | 1 | U, P | 3 | 10 | yes, models | response surface |
| (Abdul-Salam et al., 2022) | forestry | DP | uniform | 3 | to switch | 3 | OAT | 1 | OV | 1 | 1 | yes, projects | threshold diagram |
| (Guo et al., 2019) | energy | tree | BL | 1 | to expand, to abandon | 4 | OAT | 1 | OV | 2 | 3 | yes, schemes | scatter plot |
| (Ma et al., 2020) | energy | LSMC | GBM, MRP | 3 | timing, to expand | 2 | OAT | 1 | U, P | 2 | 2 | no | scatter plot, bar chart |
| (Madlener et al., 2019) | energy | tree | BL, GBM | 1 | timing | 1 | OAT | 1 | OV | 2 | 3 | yes, regimes | scatter plot |
| (Ribeiro et al., 2020) | construction | BS | GBM | 1 | whole project | 1 | OAT | 1 | P | 2 | 5 | no | no |
| (Guo & Zhang, 2020) | energy | tree | BL | 1 | to expand | 1 | GSA (Sobol) | 1 | P | 5 | 3 | no | heatmap for second-order indices, scatter plot |
| (Yao et al., 2019) | CCS | LSMC | GBM | 3 | timing, to pause, to abandon | 4 | OAT | 1 | U, OV | 4 | 3 | no | spider, bar chart |
| (Assereto & Byrne, 2021) | energy | LSMC | MRP, GARCH, GARMA | 2 | timing | 1 | OAT | 1 | OV | 2 | 2 | yes, policies | no |
| (Freitag et al., 2020) | manufacturing | tree | BL | 1 | to expand | 1 | MC | 1 | P, OV | 7 | 1 | no | histogram, scatter plot |
| (Lin & Tan, 2021) | CCS | MC | GBM, MRP | 3 | timing | 1 | OAT | 1 | U | 5 | 3 | no | bar chart, scatter plot |
| (Huang et al., 2020) | medical | tree | BL | 1 | whole project | 1 | OAT | 1 | OV | 1 | 1 | no | scatter plot |
| (Di, 2020) | energy | tree | BL | 1 | whole project | 1 | TAT | 1 | U, OV | 2 | 1 | no | no |
| (Huang et al., 2020) | energy | BS | GBM | 1 | whole project | 1 | OAT | 1 | U, OV | 4 | 1 | no | spider |
| (Polat & Battal, 2021) | transportation | BS | GBM | 1 | to expand, whole project | 2 | OAT | 1 | OV, P | 7 | 2 | no | scatter plot, bar chart |
| (Dizabadi et al., 2022) | supply chain | tree, MC | BL | 2 | whole project | 1 | OAT | 1 | U, OV | 3 | 1 | no | no |
| (Banda, 2020) | mining | fuzzy, tree | BL | 1 | to abandon | 1 | OAT | 1 | OV | 2 | 1 | yes, technologies | bar chart |
| (Seifafjei et al., 2021) | energy | LSMC, tree | BL, MRP, uniform | 3 | timing, optimal size | 2 | OAT | 1 | U, P | 3 | 3 | yes, clusters | scatter plot, bar chart |

| Paper | Application area | Valuation method | Uncertainty modelling | Number of uncertainty sources | Real option type | Number of real options | SA method | Number of methods | Type of variables for SA | Number of inputs to SA | Number of outputs for SA | Scenarios/cases | Visualization for SA |
|---|---|---|---|---|---|---|---|---|---|---|---|---|---|
| (Lee et al., 2021) | medical | Markov | Markov | 4 | option to access future treatment | 1 | OAT, MC | 2 | OV | 12 | 1 | no | tornado |
| (Li & Cao, 2022) | energy | LSMC | GBM | 2 | timing, to expand (compound) | 2 | OAT | 1 | O | 5 | 2 | no | bar chart |
| (Zhao & Min, 2020) | agriculture | DP | GBM | 1 | timing | 1 | derivatives, OAT | 2 | P | 10 | 2 | no | scatter plot |
| (Hu et al., 2022) | energy | LSMC | GBM | 3 | timing | 1 | OAT | 1 | OV, P | 6 | 2 | no | tornado, scatter plot, bar chart |
| (Xavier et al., 2020) | farming | tree | BL | 1 | to expand | 1 | NA | NA | NA | NA | NA | NA | no |
| (Najafi & Talebi, 2021) | energy | LSMC | GBM | 6 | timing | 1 | OAT | 1 | P, OV | 7 | 1 | no | scatter plot |
| (Araya et al., 2021) | mining | tree | BL | 1 | timing | 1 | GSA (Sobol) | 1 | OV | 5 | 1 | no | bar chart |
| (Liu et al., 2022) | oil&gas | LSMC | GBM | 4 | timing | 1 | OAT | 1 | OV | 6 | 2 | no | scatter plot |
| (Zeng & Chen, 2019) | energy | game, PDE | GBM | 1 | timing | 1 | OAT | 1 | OV, U, P | 7 | 3 | no | scatter plot |
| (Tolga, 2020) | medical | BS, fuzzy | GBM | 1 | whole project | 1 | derivatives | 1 | P | 4 | 1 | no | no |
| (Zhang et al., 2019) | energy | LSMC | GBM, MRP | 3 | timing | 1 | OAT | 1 | OV | 9 | 5 | no | scatter plot |
| (Durica et al., 2018) | real estate | tree | BL | 1 | to expand, to contract, to abandon | 3 | OAT | 1 | P, OV, O | 5 | 2 | no | scatter plot |
| (Mei & Wetzstein, 2017) | energy | DP | GBM | 2 | to switch | 1 | OAT | 1 | P, OV | 5 | 2 | yes, scenarios | threshold diagram |
| (Montajabiha et al., 2017) | general | BS, optimization | uniform | 1 | to abandon, compound | 1 | OAT | 1 | P | 1 | 1 | no | scatter plot |
| (Lin et al., 2019) | general | PDE | GBM | 1 | timing | 1 | TAT, OAT | 2 | P, OV | 2 | 2 | no | no |
| (Pan et al., 2019) | energy | LSMC | GBM, custom | 4 | timing | 1 | OAT | 1 | U, OV | 5 | 2 | no | scatter plot |
| (Hampe & Madlener, 2016) | energy | tree | GBM | 1 | timing, to switch | 2 | OAT | 1 | U, P, OV | 1 | 4 | no | spider, threshold diagram |
| (Power et al., 2016) | general | LSMC | GBM | 1 | other | 5 | scenario | 1 | U, P, OV | 11 | 1 | no | no |

| Paper | Application area | Valuation method | Uncertainty modelling | Number of uncertainty sources | Real option type | Number of real options | SA method | Number of methods | Type of variables for SA | Number of inputs to SA | Number of outputs for SA | Scenarios/cases | Visualization for SA |
|---|---|---|---|---|---|---|---|---|---|---|---|---|---|
| (Chladná, 2018) | medical | PDE | GBM | 1 | timing | 1 | OAT | 1 | P, OV | 2 | 3 | no | threshold diagram |
| (Kozlova et al., 2019) | energy | DP | GBM | 1 | timing, capacity choice | 2 | OAT | 1 | U, P, OV | 4 | 1 | yes, policies | scatter plot |
| (Zhang et al., 2018) | mining | PDE | MRP, GBM | 2 | timing | 1 | TAT | 1 | P, OV | 3 | 6 | no | scatter plot |
| (Rios et al., 2019) | energy | other | binomial distribution, MRP | 2 | timing | 1 | OAT | 1 | P | 2 | 1 | no | scatter plot |
| (Zhang et al., 2016) | energy | LSMC | GBM | 3 | timing | 1 | OAT | 1 | P, U, OV | 8 | 2 | no | scatter plot |
| (Zheng & Chen, 2018) | transportation | LSMC | GBM, MRP | 2 | timing | 1 | TAT | 1 | U, P | 5 | 2 | yes, policies | bar chart |
| (Delbridge & King, 2016) | agriculture | DP | MRP | 1 | timing | 1 | OAT | 1 | OV, U | 2 | 3 | yes, firm size | threshold diagram, bar chart |
| (Sim, 2018) | R&D | BS | GBM | 1 | whole project | 1 | OAT | 1 | OV | 4 | 2 | no | scatter plot |
| (Chen et al., 2018) | oil&gas | LSMC | GBM | 2 | timing | 1 | OAT | 1 | OV | 4 | 1 | yes, scenarios | scatter plot |
| (Schachter et al., 2016) | energy | MC | MRP | 1 | whole project | 1 | OAT | 1 | OV | 1 | 1 | no | scatter plot |
| (Ihm et al., 2019) | climate change | tree | scenario | 1 | timing, to abandon | 2 | OAT | 1 | U, OV | 3 | 1 | no | spider |
| (Rau & Spinler, 2017) | transportation | game, other | GBM | 1 | whole project | 1 | OAT | 1 | U, P, OV | 5 | 4 | no | box plots |
| (Moon & Baran, 2018) | energy | DP | GBM | 1 | timing | 1 | OAT, TAT | 2 | OV | 1 | 2 | no | scatter plot, response surface |
| (Adkins & Paxson, 2017) | general | PDE | GBM | 1 | to abandon | 1 | OAT | 1 | P | 3 | 2 | no | scatter plot |
| (García-Ramos et al., 2016) | farming | BS | GBM | 1 | whole project | 1 | OAT | 1 | P | 2 | 1 | no | no |
| (Luo & Kou, 2018) | transportation | PDE | MRP | 1 | timing | 1 | OAT | 1 | P | 6 | 2 | no | scatter plot |
| (Abadie et al., 2017) | construction | MC | GBM | 1 | other | 1 | OAT | 1 | P | 2 | 3 | yes, processes | no |
| (Kim et al., 2017) | metallurgy | BS | GBM | 1 | whole project | 1 | TAT | 1 | P | 2 | 1 | no | response surface |

| Paper | Application area | Valuation method | Uncertainty modelling | Number of uncertainty sources | Real option type | Number of real options | SA method | Number of methods | Type of variables for SA | Number of inputs to SA | Number of outputs for SA | Scenarios/cases | Visualization for SA |
|---|---|---|---|---|---|---|---|---|---|---|---|---|---|
| (Kauffman et al., 2015) | IT | LSMC | MRP | 2 | timing | 1 | OAT | 1 | P | 2 | 2 | no | histogram |
| (Azimi et al., 2013) | mining | LSMC | GBM | 2 | timing | 1 | OAT | 1 | P | 3 | 1 | no | scatter plot |
| (Wu et al., 2015) | real estate | BS | GBM | 1 | compound, whole project | 2 | OAT | 1 | P | 8 | 1 | no | scatter plot |
| (Cardin et al., 2015) | oil&gas | MC | normal | 2 | whole project | 1 | TAT, OAT | 2 | P | 5 | 2 | no | scatter plot, tornado |
| (Shen & Pretorius, 2013) | real estate | tree | BL | 1 | whole project | 1 | OAT | 1 | P | 4 | 1 | no | no |
| (Kruk & Nikulina, 2016) | oil&gas | tree | BL | 1 | to expand | 1 | OAT | 1 | OV | 3 | 1 | no | spider |
| (Armada et al., 2013) | general | PDE | GBM | 2 | whole project | 1 | derivatives | 1 | P | 3 | 3 | no | scatter plot |
| (Pringles et al., 2015) | energy | LSMC | GBM, MRP, Markov | 3 | timing | 1 | OAT | 1 | P | 4 | 1 | no | scatter plot |
| (Bosch-Badia et al., 2015) | general | BS | GBM | 1 | whole project | 1 | TAT | 1 | P | 4 | 1 | yes | scatter plot, response surface |
| (Guthrie, 2012) | general | PDE | GBM | 1 | timing, to expand | 2 | TAT | 1 | P | 3 | 4 | no | scatter plot |
| (Klepáč et al., 2013) | general | BS, tree, MC | GBM, jump | 1 | to expand, to abandon | 2 | OAT | 1 | OV, P | 6 | 3 | no | spider |
| (Mirkhani & Saboohi, 2012) | energy | tree | BL | 1 | timing | 1 | OAT | 1 | P | 2 | 1 | no | no |
| (MacDougall, 2015) | energy | BS | GBM | 1 | timing | 1 | OAT | 1 | P, OV | 3 | 1 | no | spider |
| (Huang et al., 2014) | mining | PDE | GBM | 2 | timing | 1 | TAT | 1 | P, OV, U | 5 | 1 | no | response surface |
| (Chen, 2012) | supply chain | PDE | GBM | 1 | timing | 1 | TAT | 1 | P, OV, U | 12 | 3 | no | response surface |
| (Jeuland & Whittington, 2014) | water | MC | scenario | 2 | whole project | 1 | OAT | 1 | U | 2 | 1 | yes, scenarios, areas | spider |
| (Deng et al., 2013) | water | MC | GBM, scenario | 1 | custom | 1 | OAT | 1 | OV | 9 | 2 | no | spider |

| Paper | Application area | Valuation method | Uncertainty modelling | Number of uncertainty sources | Real option type | Number of real options | SA method | Number of methods | Type of variables for SA | Number of inputs to SA | Number of outputs for SA | Scenarios/cases | Visualization for SA |
|---|---|---|---|---|---|---|---|---|---|---|---|---|---|
| (Yin & Hu, 2007) | general | game, PDE | GBM | 1 | timing | 1 | derivatives | 1 | OV | 2 | 1 | no | no |
| (Park & Lim, 2009) | climate change | PDE | GBM | 1 | timing | 1 | TAT | 1 | P | 2 | 1 | yes, plants | response surface |
| (Abdel et al., 2008) | mining | LSMC | GBM, MRP | 3 | timing | 1 | OAT | 1 | P | 1 | 1 | no | no |
| (Park, 2008) | IT | tree | BL | 1 | timing | 1 | OAT | 1 | P | 4 | 1 | no | no |
| (Costantino & Pellegrino, 2010) | supply chain | MC | custom, BL | 2 | other | 1 | TAT | 1 | P | 3 | 1 | no | histogram |
| (Çagri et al., 2010) | customer service | BS, tree, fuzzy | trapezoidal fuzzy number | 2 | whole project | 1 | TAT | 1 | P | 2 | 1 | yes, methods | response surface |
| (Harmantzis & Tanguturi, 2007) | telecom | BS | GBM | 1 | timing | 1 | OAT | 1 | P, U | 3 | 2 | yes, cases | scatter plot |
| (Rammerstorfer & Eisl, 2011) | CCS | PDE | MRP, ABM | 3 | timing | 1 | OAT | 1 | P, OV | 10 | 2 | no | scatter plot |
| (Wang & Li, 2010) | agriculture | MC | MRP | 2 | whole project | 1 | TAT, OAT | 2 | P | 7 | 2 | no | scatter plot |
| (Lee et al., 2008) | general | other | GBM | 1 | compound | 1 | derivatives | 1 | P | 2 | 2 | no | no |
| (Camargo et al., 2011) | transportation | BS | GBM | 2 | whole project | 1 | TAT | 1 | P, OV | 4 | 1 | no | scatter plot |
| (Wu, 2011) | patents | BS | GBM | 1 | whole project | 1 | derivatives | 1 | P | 4 | 1 | no | no |
| (Tseng & Lin, 2007) | energy | tree | BL | 2 | other | 1 | OAT | 1 | P | 3 | 1 | no | scatter plot |
| (Reindorp & Fu, 2011) | general | PDE | GBM | 2 | timing | 1 | TAT | 1 | P | 2 | 2 | no | no |
| (Wu & Yen, 2007) | R&D | BS | jump, GBM | 1 | to expand | 1 | TAT | 1 | P | 8 | 1 | no | scatter plot |
| (Chladná, 2007) | forestry | PDE | MRP | 2 | timing | 1 | TAT | 1 | P, OV, U | 5 | 1 | no | threshold diagram |
| (Sarkar, 2009) | farming | PDE | stochastic logistic growth | 1 | timing | 1 | OAT | 1 | P | 4 | 3 | no | no |
| (Anderson et al., 2004) | general | PDE | GBM | 1 | other | 1 | TAT | 1 | P, U | 2 | 1 | no | scatter plot |

| Paper | Application area | Valuation method | Uncertainty modelling | Number of uncertainty sources | Real option type | Number of real options | SA method | Number of methods | Type of variables for SA | Number of inputs to SA | Number of outputs for SA | Scenarios/cases | Visualization for SA |
|---|---|---|---|---|---|---|---|---|---|---|---|---|---|
| (Bendall & Stent, 2005) | transportation | MC, tree | triangular | 4 | to expand, other | 2 | scenario | 1 | P | 4 | 1 | no | no |
| (Nembhard et al., 2003) | supply chain | MC | GBM | 3 | whole project | 1 | OAT | 1 | P | 5 | 1 | no | scatter plot, tornado |
| (Schwartz & Moon, 2000) | telecom | PDE | GBM, MRP | 1 | whole project | 1 | OAT | 1 | P | 15 | 3 | no | no |
| (Miller & Bertus, 2005) | transportation | tree | BL | 1 | whole project | 1 | OAT | 1 | P | 4 | 1 | yes, licences | spider |
| (Murillas & Chamorro, 2006) | farming | PDE | GBM, stochastic logistic growth | 1 | timing | 1 | TAT | 1 | U | 2 | 2 | yes, scenarios | scatter plot |
| (Saphores, 2000) | agriculture | PDE | diffusion | 1 | timing | 1 | OAT, TAT | 2 | P | 11 | 3 | no | scatter plot |
| (Lewis et al., 2004) | R&D | BS | GBM | 1 | timing | 1 | OAT | 1 | P, OV | 5 | 1 | no | spider |
| (Liukkonen et al., 2025) | energy | MC | uniform, triangular | 4 | whole project | 1 | GSA (simdec) | 1 | U, OV | 6 | 1 | yes, cases | simdec, histogram |